\documentclass[sigconf]{acmart}

\AtBeginDocument{%
  \providecommand\BibTeX{{%
    \normalfont B\kern-0.5em{\scshape i\kern-0.25em b}\kern-0.8em\TeX}}}

\copyrightyear{2025}
\acmYear{2025}
\setcopyright{acmlicensed}\acmConference[CHI '25]{CHI Conference on Human Factors in Computing Systems}{April 26-May 1, 2025}{Yokohama, Japan}
\acmBooktitle{CHI Conference on Human Factors in Computing Systems (CHI '25), April 26-May 1, 2025, Yokohama, Japan}
\acmDOI{10.1145/3706598.3713232}
\acmISBN{979-8-4007-1394-1/25/04}

\newenvironment{myquote}%
  {\list{}{\leftmargin=0.1in\rightmargin=0.1in}\item[]}%
  {\endlist}

\begin{document}

\title{The Datafication of Care in Public Homelessness Services}


\author{Erina Seh-Young Moon}
\affiliation{%
\institution{University of Toronto}
  \city{Toronto}
  \country{Canada}}
\email{erina.moon@mail.utoronto.ca}
\orcid{0000-0003-3233-9773}

\author{Devansh Saxena}
\affiliation{%
\institution{University of Wisconsin-Madison}
  \city{Madison}
  \country{United States}}
\email{devansh.saxena@wisc.edu}
\orcid{0000-0001-5566-7409}

\author{Dipto Das}
\affiliation{%
\institution{University of Toronto}
  \city{Toronto}
  \country{Canada}
 }
\email{diptodas@cs.toronto.edu}
\orcid{0000-0001-5704-8804}

\author{Shion Guha}
\affiliation{%
\institution{University of Toronto}
  \city{Toronto}
  \country{Canada}
 }
\email{shion.guha@utoronto.ca}
\orcid{0000-0003-0073-2378}

\renewcommand{\shortauthors}{Moon et al.}

\begin{abstract}
Homelessness systems in North America adopt coordinated data-driven approaches to efficiently match support services to clients based on their assessed needs and available resources. AI tools are increasingly being implemented to allocate resources, reduce costs and predict risks in this space. In this study, we conducted an ethnographic case study on the City of Toronto’s homelessness system’s data practices across different critical points. We show how the City’s data practices offer standardized processes for client care but frontline workers also engage in heuristic decision-making in their work to navigate uncertainties, client resistance to sharing information, and resource constraints. From these findings, we show the temporality of client data which constrain the validity of predictive AI models. Additionally, we highlight how the City adopts an iterative and holistic client assessment approach which contrasts to commonly used risk assessment tools in homelessness, providing future directions to design holistic decision-making tools for homelessness. 
\end{abstract}

\ccsdesc[500]{Human-centered computing~Empirical studies in HCI}
\ccsdesc[500]{Applied computing~Computing in government}

\keywords{algorithmic decision-making, algorithmic bias, risk assessments, homelessness, public sector}


\maketitle

\section{Introduction}

Public sector agencies have long used data as the basic building blocks to drive welfare policies, using data to identify the types of clients they serve and providing services to specific population subgroups that are deemed `deserving’ of scarce social assistance resources through evidence-based, defensible decision-making \cite{Levy_2021, johnson22, eubanks2018automating, redden2020, kelly23, nielsen2023, saxena2020human}. The role of data has become increasingly critical as more agencies are moving from categorical prioritization approaches which allocate services based on eligibility criteria to algorithmic prioritization approaches using artificial intelligence (AI) decision-support tools \cite{johnson22, sambasivan21}. In fact, AI systems' failures, validity concerns, and unfair outcomes to downstream users are often attributed to issues around data representativeness, quality, and completeness \cite{raji22, suresh21, Reinmund2024}. And yet, despite data’s touted importance, AI developers consistently under-value data work by making assumptions about the data-generating process and validating models using goodness-of-fit metrics that are removed from the situatedness of the domain \cite{sambasivan21, raji22, selbst2019fairness}.

In this paper, we examine the data practices of frontline staff within a large homelessness support system in a major Canadian city, Toronto, Canada. Over the last two decades, North American governments have mandated the implementation of coordinated data-driven approaches to streamline support services to people who are at risk of or are experiencing homelessness (also known as \textbf{“coordinated systems”}) \cite{reachinghomes, openingdoors, Ecker2022}. Moreover, following explosive interest in AI, cities and states in Canada and the US are increasingly turning to data and AI to track clients and predict their risk of experiencing homelessness \cite{kithulgoda_predictive_2022, ottawa, laai_2024article, ny_track, vanberlo2020}. As a notable example, in the US, the state of California recently issued an open call to developers to submit proposals on how generative AI can be used to combat the homelessness crisis \cite{cali_genai}. Moreover, in Canada, Ottawa and London are turning to AI to predict chronic homelessness \cite{ottawa, vanberlo2020}. In line with these trends, the SIGCHI community has also made significant strides to study applications of data and AI in homelessness. Kuo et al. \cite{kuo23} elicited stakeholder perceptions around housing prioritization AI tools, Showkat et al. \cite{showkat23} examined values in algorithms for homelessness \cite{showkat23}, and Moon and Guha \cite{moon24} unpacked the technical underpinnings of these tools. Through these works, researchers have raised concerns about these tools’ validity and use of decontextualized, easy-to-quantify administrative data. Furthermore, Slota et al. \cite{slota23} and Karusala et al. \cite{karusala19} highlight the critical need to understand frontline workers’ data practices because value tensions arising from data work significantly impact the accuracy and consistency of client data public agencies collect.

Motivated by the growing adoption of AI tools in homelessness and the critical role of data, our study builds on prior studies on stakeholders' reported homelessness data practices \cite{slota23, karusala19}. In our work, we  take a \textit{systems-approach} \cite{kelly24, saxena2021framework2} to study interconnected frontline worker data practices through interviews and observations. In close partnership with the City of Toronto that experiences great demand for shelter and housing services, we conducted an in-depth ethnographic case study to understand how workers engage in differential data practices at different critical points of the homelessness system \cite{toronto}. Over 3 months, we conducted semi-structured interviews with 31 staff, and observed 21 staff for around 60 hours at 8 different service provider groups within the City to unpack \textit{reported} data practices of frontline workers and \textit{observe} their not-vocalized data work. Through this work, we sought to uncover how interactions between organizational data requirements, resource constraints, and (at times) resistance from clients shape how frontline staff collect homelessness client data and its greater ramifications on AI research for homelessness. Our study thus asks the following research questions: 

\begin {itemize} 

\item \textbf{RQ1:} \textit{How do public homelessness systems collect client information?}

\item \textbf{RQ2:} \textit{How do public homelessness systems use data to support clients through the homelessness system?}

\item \textbf{RQ3:} \textit{How can HCI researchers support data-driven homeless support systems and frontline staff?}
\end{itemize}

This paper makes the below unique research contributions: 

\begin {itemize} 
\item We take a systems-approach \cite{saxena2021framework2, kelly24} to investigate how frontline staff engage in differential data practices across different service providers within the homelessness system. We highlight how homelessness system’s data practices provide standardized processes for workers to navigate uncertainty and provide care to clients, but technological, spatial, and staffing-related factors can constrain intended data practices across service provider sites. 
\item We showcase how workers engage in heuristic decision-making, navigating different and, at times, conflicting care-driven data collection objectives to help clients access critical services they need.
\item We illustrate how worker’s heuristic decision-making result in data practices that provide care \cite{mol2010} for clients but also complicates how AI developers can use homelessness data for model building.
\item We reveal how holistic client needs assessments can be implemented in the public sector, signaling how HCI researchers can shift away from predictive risk models for homelessness \cite{kithulgoda_predictive_2022, kuo23, toros18}. We also surface current limitations to implementing this holistic approach. 

\end{itemize}

This work responds to calls within SIGCHI research to investigate the data practices of high-stakes sociotechnical systems where the deployment of AI can have disproportionately negative effects on vulnerable communities \cite{moon24, sambasivan21, showkat23}. In the next section, we provide an overview of related work that have motivated our study's RQs.

\section{Related Work}

\subsection{SIGCHI research on public sector data}

Public sector agencies in North America have a long history of collecting data on individuals, but as governments turn to data-centric practices, greater value is ascribed to public sector data with the expectation that it can be transformed into actionable insights and guide evidence-based decision-making \cite{Dencik_2019, Levy_2021}. In line with this movement, the SIGCHI community has actively studied public sector sociotechnical systems, examining values and stakeholder perspectives around public technologies \cite{haque2024we, kuo23, kelly24, kawakami2022, saxena2021framework2, stapleton2022has}, engaging in participatory design work that uplift and empower vulnerable communities \cite{LeDantec_2011, LeDantec2009, Yoo_2013, gondimalla24, Holten2021}, and pushing the boundaries of HCI methodologies to deeply engage with the public sector \cite{kuo23, halperin23, haque2024we, saxena2020conducting, wan2023community}. Voida et al. \cite{Voida2014} and Møller et al. \cite{Holtenshifting_2020} highlight digital technologies offer affordances to increase governmental efficiencies, expand access to services, and greater opportunities to disseminate information. And yet, conflicting logics can emerge where important questions on whose values should drive the design of public technologies, how, and if those values can be operationalized \cite{Voida2014}. 

Recently, SIGCHI research has turned their focus on studying algorithmic or AI decision-support systems in the public domain as public agencies increasingly pull data from various public data sources - including data from homelessness management information systems (HMIS), social assistance programs, health services, criminal records, child welfare services, and more - to extract client information, identify common patterns, predict client outcomes, and allocate services to clients \cite{kithulgoda_predictive_2022, afst_documentation, eubanks2018automating, toros18}. While these data-driven systems were created in the hopes that big data can help pre-emptively target resources or interventions consistently for those in need \cite{Dencik_2019}, a plethora of SIGCHI work has found these tools often reductively conceptualize client risk \cite{brown19, saxena2020human, moongi24}, and result in biased outcomes for vulnerable population groups \cite{ feng22, stapleton22}. Veale et al. \cite{veale18} argue many of these issues arise because the tools are being developed in isolation from their specific context. In turn, Sambasivan et al. \cite{sambasivan21} and Suresh and Guttag \cite{suresh21} cite the importance of data work, highlighting that poor data quality and weak incentives to ensure data excellence (i.e., the practice of diligent data documentation, sustained partnerships with data domain experts) can cause algorithmic or AI harms. In light of growing awareness of harms that may arise from such public technologies, HCI scholars have emphasized the importance of engaging in participatory and collaborative work with stakeholders and the need to deeply understand domain-specific data practices when developing tools using public sector data \cite{saxena2021framework2, saxena2020human, kawakami2022, kuo23, showkat23}. 

\subsection{Data and care work}

Accompanying the greater role of data in public agencies, SIGCHI literature has long interrogated the construct of data highlighting its situatedness \cite{vertesi2011, Bopp_Harmon_Voida_2017, Koesten_2019, Feinberg_2017, Muller_2019_howdsworks, pine2015, chen_covid23, kapania_data23}. Notably, Vertesi and Dourish \cite{vertesi2011} showed how organizational culture can influence data collection practices where \textit{"work with data enacts social relationships"}. Similarly, Bopp et al. \cite{Bopp_Harmon_Voida_2017} investigating data practices in mission-driven organizations have highlighted how conflicting stakeholder interests can influence what metrics an organization collects and can inadvertently erode organizational autonomy, subverting naïve expectations that data can always improve organizational practices. Because most public agencies are structures that support citizen welfare \cite{Light2019}, SIGCHI research on public data work has been closely tied with relational and democratic notions of care \cite{Gilligan_2003, Tronto_2013, mol2010}. Studies in this area have found public sector workers are primarily motivated to provide client care but must also juggle conflicting stakeholder interests and technological constraints, which result in contextualized data practices \cite{boone2023, tran2022, nielsen2023, karusala19, saxena2021framework2, Holtenshifting_2020, slota23}. For example, Tran et al. \cite{tran2022} showed non-profit workers who prioritized care for clients engaged in fragmented data practices through an assemblage of homebrewed ICT tools. Additionally, Boone et al. \cite{boone2023} showed how a food assistance program's lean data practices were shaped by negotiating between different factors: the need to provide financial records for grant fund applications and audits while protecting clients' immigration status and providing low barrier access to services. Studies have also shown that caring practices in public sociotechnical systems are often relational, where trust and rapport are critical factors for data production \cite{karusala19, nielsen2023}. For example, Nielsen et al. \cite{nielsen2023} revealed that caseworker's attentiveness and rapport with clients can nudge clients to share personal information, which caseworkers would then translate into relevant and credible data. Homelessness systems are designed to care for unhoused clients, and as such, following current literature on public data, our study was interested in understanding if and how care is enacted through its data practices.

\subsection{HCI research on homelessness}

In recent years, to alleviate the growing demand for shelter and housing services, communities in North America have begun implementing coordinated data-driven approaches using standardized client assessments to assess a client's risk of homelessness and prioritize scarce services to them \cite{Ecker2022, reachinghomes, openingdoors} (see Section \ref{sec:researchcontext} for more information). This approach takes on a system-level approach to supporting clients where client information is tracked on a real-time basis and shared among different service providers to minimize repeated information collection and encourage collaborative client support \cite{Slota_2021, Ecker2022}. In response, recent HCI work has turned their efforts to examine data practices within homelessness systems by interviewing stakeholders \cite{karusala19, slota23, gondimalla24, Slota_2021, karusala_contestability24, Slota_Fleischmann_Greenberg_2022, Slota_infra2022,Slota_caring_2023}. For example, Slota et al. \cite{Slota_infra2022} showed that although homeless systems' data infrastructures enable data sharing between service providers, workers find client information that is being shared is not always sufficient or usable. Karusala et al. \cite{karusala19} also found client data did not necessarily guide how clients are matched to services; instead, a caseworker's assessment of a client's vulnerability would be formalized \text{into} data. For example, if there was a mismatch between a client's risk assessment score and a worker's perception of their vulnerability, workers would sometimes reassess clients to justify matching a client to a particular service. Tracey and Garcia \cite{Tracey_Garcia_2024_automation} also revealed data documentation requirements can detract workers from providing direct care to clients. Additionally, given the increasing pervasiveness of AI decision-support tools and risk assessment algorithms in homelessness service delivery decisions, the HCI community has critically examined these tools \cite{Tracey_Garcia_2024_automation, Slota_infra2022, Slota_caring_2023}. For example, Slota et al. \cite{slota23} problematized using commonly used risk assessments such as the Vulnerability Index-Service Prioritization Decision Assistance Tool (VI-SPDAT) tool as a prioritization tool in homelessness systems because it relies on self-reported client data but very vulnerable clients cannot always advocate for themselves. Moreover, Moon and Guha \cite{moon24} and Showkat et al. \cite{showkat23} found such tools are often built using messy and biased public datasets that ignore human values and unduly focus on profiling the client's risk level without fully accounting for how the resource-constrained system can exacerbate client vulnerabilities.

Prior work to date on coordinated data-driven homelessness systems has conducted interview studies \cite{slota23, Slota_2021, karusala19, Tracey_Garcia_2024_automation, Slota_Fleischmann_Greenberg_2022, Slota_caring_2023, Slota_infra2022} and comic boarding workshops \cite{kuo23, gondimalla24} to understand stakeholders' \textit{reported} tensions and values that emerge in these sociotechnical systems. Our study sought to build on these works further by conducting interviews with- and observations of- frontline staff working at different critical points of a homelessness support system to gain a systems-level understanding \cite{kelly24, saxena2021framework2} of both reported and observed worker data practices. Our study's research questions arose after several meetings with the City of Toronto that faces high demand for shelter and housing assistance. City staff expressed the desire to holistically understand how workers engage in similar or different data practices \textit{across their system} to improve their overall data practices. Through our in-depth ethnographic study, we discovered workers engaged in differential data practices depending on the service provider’s role within the overall homelessness system as they navigated uncertainty, resource constraints, and mandated data procedures, all while prioritizing care to clients.

\section{Research Context} \label{sec:researchcontext}

\subsection{Coordinated entry and access systems in the US and Canada}

In the face of rising homeless populations, many North American homelessness support systems are comprised of a complex network of governmental and non-governmental agencies that provide clients with critical supports (i.e., emergency shelter, housing, medical or educational supports, counseling) \cite{Fowler_system_2022}. To ensure these organizations are acting collectively in a way that is transparent and efficient rather than on an ad-hoc, first-come-first-serve basis, North American federal governments have mandated the adoption of coordinated entry and coordinated access systems for homelessness, which we collectively term as \textbf{coordinated systems} in this paper \cite{reachinghomes, openingdoors, Ecker2022}. Through this approach, the US Department of Housing and Urban Development and the Canadian Reaching Home Directives aim to streamline how clients are referred to services and encourage information sharing between service providers to allocate services efficiently and minimize repeated information collection. Coordinated systems, at a minimum must have \textbf{1)} have clear entry points for individuals to access a homelessness support system (e.g., through a central call center). \textbf{2)} Communities must use standardized assessments to assess client needs and vulnerability. Many communities use algorithmic risk assessment tools such as the VI-SPDAT, and increasingly, cities are turning to AI decision-support tools \cite{Ecker2022, showkat23, moon24}. However, prior work has found these tools are often biased against historically marginalized communities\footnote{In fact, the creators of the popular VI-SPDAT tool stated they no longer support the tool \cite{orgcode}} \cite{shinn22, brown_2018}. Following client assessments, \textbf{3)} clients are to be prioritized for services, and finally, \textbf{4)} prioritized clients are matched to available supports. Acknowledging that communities face vastly different local challenges, communities have considerable freedom to decide what assessment tools to use and how to prioritize clients \cite{Ecker2022, coordinatedguide, reachinghomesdirectives, hudselfassess}. The implementation of coordinated systems places greater importance on the role of data for homelessness support, and communities often use a centralized database to implement coordinated systems, the Homeless Management Information System (HMIS), to track clients seeking services, their service history within a homelessness system, and available supports within the system \cite{showkat23, hifis, hudhmis}. On the immediate level, HMIS data is intended to identify and match clients with appropriate services within coordinated systems. At the same time, aggregated data collected through HMIS are used to understand temporal homelessness patterns and directly guide policy on which clients should be prioritized for scarce shelter space/housing resources and determine how much housing-related services a community needs in the long-term \cite{systems_NAEH23, canada_homelesseval_2018}. Despite these intended uses, significant on-the-ground realities complicate intended data practices. People experiencing homelessness often deeply mistrust workers and lack confidence in homeless service providers because of previous negative experiences where they have been treated with apathy, disrespected, faced unsafe shelter conditions, and been repeatedly let down by frontline workers or the system that only provides short-term solutions \cite{Kryda_Compton_2009, Hoffman_coffey_dignity, Welsh_blacklives, karusala19}. When client mistrust is combined with a homelessness system that asks clients lengthy and intrusive (potentially trauma-triggering questions), clients are unwilling to share their information \cite{Welsh_blacklives}. 

\subsection{Site of research inquiry}
\begin{figure*}[]
\centering 
\includegraphics[scale=0.25]{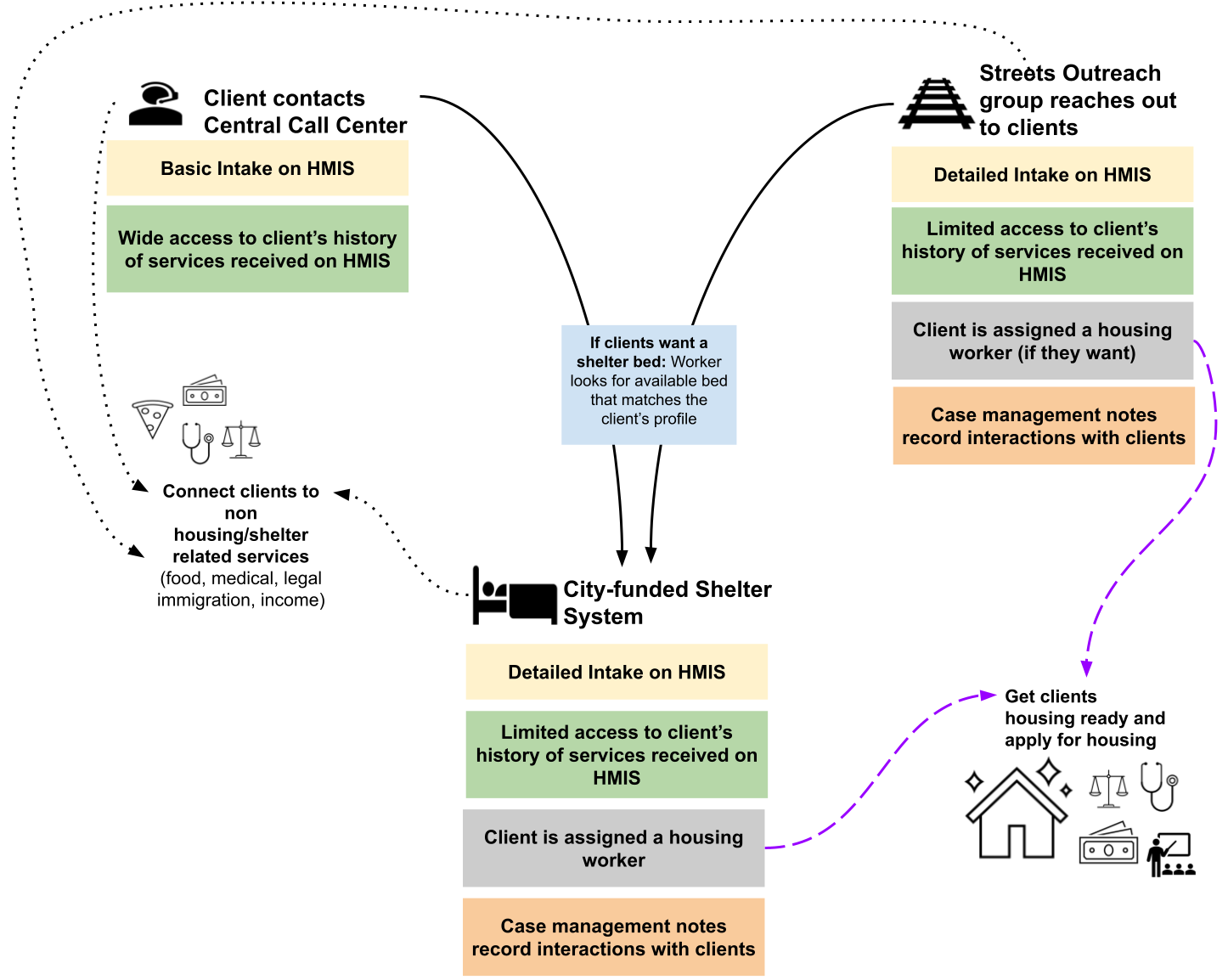}
\caption{Mapping of the City's shelter system and data practices. (This is a simplified diagram of the City's homelessness system used to illustrate the coordination efforts between different groups and each groups's data practices)}
\label{fig:system}
\end{figure*}

In this study, we partnered with the City of Toronto, a large Canadian city that is experiencing high demand for shelter and housing assistance. The City has developed its own unique client assessment tool - the STARS (Service Triage, Assessment, and Referral Support) common assessment tool - in line with coordinated system requirements set by the Canadian government to provide standardized processes for frontline staff to understand people’s needs and assist them \cite{reachinghomesdirectives, coordinated_toronto}. The tool consists of three components: 1) an intake and triage process, 2) tracking whether clients have documents needed to apply for housing, and 3) a housing assessment tool to determine what follow-up supports a client needs once housed. The first and second components of the tool are implemented within the City’s proprietary HMIS called the Shelter Management Information System (SMIS). The first component is an important step for the City because here, workers collect basic client information to 1) identify the immediate supports they need so workers can direct them to available resources (e.g., harms reduction services, apply for ID); and 2) help workers determine their caseload. Following federal guidelines, the first component also helps the City understand, on a macro level, the clientele they serve, their needs, outcomes, and service and system disparities to set homelessness reduction targets and guide long-term planning on how to improve their services for different client groups (e.g., Indigenous and chronic homelessness) \cite{reachinghomesdirectives}. Client intakes occur at different points of the homelessness system - primarily when a client first accesses services at a different service provider and intake information is continuously updated as clients share new information with workers. The City expressed an interest in identifying best practices that support the collection of accurate and consistent data on their HMIS database during the intake process across its expansive homelessness service network. Through discussions with the City, we learned the City has three different service provider groups where critical client intakes occur: a central call center, streets outreach group, and shelters. Our study therefore focused on examining data practices of front-line workers during the first component of the City’s assessment tool at these sites. Figure \ref{fig:system} shows how clients typically move from one service provider type to the next, and in the following paragraphs, we briefly detail the intake process for each of these service groups informed by our ethnography and conversations with City staff.

\textbf{Central Call Center (‘Call Center’)} In line with the City's coordinated access approach, which requires having clear and consistent access points into the homelessness system \cite{reachinghomesdirectives, Ecker2022}, clients seeking homelessness-related services such as detox programs, local food banks, and shelter are directed to call a 24/7 hotline center. If the client is calling the Call Center for the first time for a shelter bed, a caseworker will conduct a brief `Intake' by asking a series of basic questions, including their name, gender, pronouns, and accessibility needs. They will then save their information on the City's HMIS and issue the client an HMIS number. Each shelter bed offered within the City's shelter system is reserved for specific types of clients. For example, a bed may be reserved for women or those who require accessibility support. Based on client `intake' data, the Call Center worker will search for an available shelter bed on their HMIS that matches the client's profile and needs. If a shelter bed that matches the client's profile is available at the time of the call and they are not restricted from staying at the shelter, the worker will offer them the space. If no spaces are available when the client calls into the Call Center (which is often the case), the client will be asked to call again in the next few hours so the worker can see if shelter beds have opened up matching the client's profile. 

\textbf {Streets Outreach Group (SOG)} The City has a SOG that primarily conducts street outreach and offers housing-related support to clients experiencing street homelessness. SOG workers go out daily to subways, alleys, and encampments to regularly engage with clients, conduct wellness checks, and help clients put together documentation to apply for housing as they continue to live outdoors. If a client expresses interest in entering a shelter or housing, SOG workers will complete a client intake similar to the process at the Call Center but ask more detailed questions, such as their physical and mental health needs. SOG workers can help clients find shelter spaces by calling the Call Center on their behalf or contacting their team leader to see if there are reserved shelter beds for SOG clients. Following a client’s intake on HMIS, SOG workers record subsequent interactions with the client through casenotes on HMIS and update the client's intake form as any new client information emerges. 

\textbf{Shelters} The City offers various overnight services for clients experiencing homelessness, including shelters for families, single men or women, youth, and emergency overnight programs during extreme weather conditions. For this paper, we will refer to all these service providers as \textit{shelters}. When a client is referred to a shelter by the \textit{Call Center} or the \textit{Streets Outreach Group} and first enters the shelter, they are required to complete a new `intake’ on HMIS. Frontline staff ask clients questions similar to the intake at the Call Center but also ask detailed questions, including the client’s reason for homelessness, employment status, physical or mental health history, substance use problems, allergies, and dietary restrictions (since shelters provide meals). Shelter workers record subsequent interactions with the client through narrative casenotes on HMIS and update the client's intake form if a client shares new or updated information.

\section{Methods}

This study's research objectives emerged after multiple meetings with a supervisor in the City. We discussed the City's current data practices and how we can help improve the City's data needs. Through conversations, we collaboratively decided that an ethnography on the City's intake processes could help the City identify best practices to improve its data collection processes. Client intake information is conducted through and saved on the City's HMIS as part of the first component of the client assessment tool (see Section \ref{sec:researchcontext}). It is a critical step in supporting clients because the information can help workers quickly identify client needs and help drive systemic change to the City's homeless service provision. Due to the important role of client intake, the City was interested in how they can improve their intake processes. Before beginning our study, we obtained approval from the Research Ethics Board (REB) from our research institution.

\begin{table*}[]

\begin{tabular}{lccc}
\multicolumn{1}{l|}{\textbf{Site types visited}} & \multicolumn{1}{c|}{\textbf{Client support worker}} & \multicolumn{1}{c|}{\textbf{Housing worker}} & \textbf{Supervisor} \\ \hline
\multicolumn{1}{l|}{Outdoors street outreach group} & \multicolumn{1}{c|}{P40, P1, P2, P1111, P8789} & \multicolumn{1}{c|}{P39} &  \\
\multicolumn{1}{l|}{Central call center} & \multicolumn{1}{c|}{P300, P3978, P111} & \multicolumn{1}{c|}{} & P888 \\
\multicolumn{1}{l|}{Partner-run shelters} & \multicolumn{1}{c|}{P44, P45, P46, P47, P990, P991, P677, P678, P679} & \multicolumn{1}{c|}{P76} & P25 \\
\multicolumn{1}{l|}{City-run shelters} & \multicolumn{1}{c|}{P35410, P35420, P867, P211, P222, P987, P9985} & \multicolumn{1}{c|}{P88} & P789, P349, P9277, P82 \\
\multicolumn{1}{l|}{} & \multicolumn{1}{c|}{P7354, P3, P31} & \multicolumn{1}{c|}{} &  \\
 & \multicolumn{1}{l}{} & \multicolumn{1}{l}{} & \multicolumn{1}{l}{}
\end{tabular}%

\caption{Sites visited and study participant role. Note: Partner-run shelter are shelters run by external organizations funded by the city. We do not include exact participant titles for privacy reasons. }
\label{tab:participant}
\end{table*}


\textbf{Recruitment \hspace{0.3cm}} This study adopted a purposive sampling approach following several steps. Because our study's goal was to gain a system-level understanding of the City's intake processes, we first worked with the City to identify which critical service provider groups we wanted to recruit participants from. Through discussions, we determined that we should recruit participants from the City's central call center, which acts as the main access point for clients seeking services, the City's streets outreach group, which conducts streets outreach, and shelters operated by the City and partner organizations (partner organizations are funded by the City but are run by external organizations). The City helped us contact supervisors for the different service provider groups. The first author then met with each supervisor to explain the study's aims. Following the meetings, interview and observation participants were recruited in various ways. Some supervisors found workers in their group willing to participate in the study; other times, the first author was invited to come to the site, talk to workers about the study, and find workers willing to participate. Through these varied approaches, the first author visited 8 different service providers on 19 occasions over 3 months; conducting semi-structured interviews with 31 staff, and observing 21 staff for around 60 hours. In Table \ref{tab:participant}, we present an overview of the different sites and participants we interviewed and observed.

\textbf{Interviews and Observations\hspace{0.3cm}} Interviews and observations took place during workers' work hours. It was not always possible to record interviews with participants because some participants did not want to be recorded, and some interviews took place in open, bustling spaces. In these cases, the first author took copious notes. Observations took many forms. In some instances, the first author was invited to follow a worker's entire shift to observe their data work throughout the day. In other cases, the first author sat with workers for several hours to observe their work. With consent, the first author also observed workers conduct client intakes over the phone or in person. During observations, the first author asked follow-up questions when workers had time to answer questions.

\textbf{Data Analysis \hspace{0.3cm}} The first author transcribed recorded interviews verbatim. Notes the first author took if the interviews were not recorded and observational notes were compiled into a debriefing document after the site visit on the same day. After each site visit, the first author shared their immediate findings with the last author and regularly met with co-authors to discuss and synthesize findings. We used new insights that emerged from our field work to probe deeper into our high-level research questions. To analyze our data and answer our RQs, we applied thematic qualitative analysis \cite{braunclark2006}. The first author read and reread interview transcripts and field notes numerous times to create initial codes and map the codes visually. Then, the first author consulted with other co-authors to reach a consensus on the codes, resolve ambiguities, and conceptually group them into themes. We also shared our initial findings with City supervisors to further contextualize our findings. 

In the following section, we detail our study findings. Given the intense public scrutiny of homeless support services in the City, we have taken extra precautions to ensure participant privacy. We do not provide detailed demographic or specific title breakdowns of the study participants and omit participant IDs from potentially sensitive quotes.

\section{Results}

\subsection{Care-driven objectives shape HMIS data practices} \label{sec:r2}

Our study's deep interrogation into the City's data practices revealed that its practices were centered around `care,’ which, following Mol et al. \cite{mol2010}, relates to “a negotiation about how different goods might coexist in a given, specific, local practice.” (p. 13). \textbf{Our findings showed that client care was datafied through the City’s data practices, providing standardized processes for workers to navigate the inherent uncertainties that underlie client support.} Through our study, we identified three (at times competing) care-driven objectives in our examination of the City's data practices, which at times worked in tandem but also against each other. Below, we describe how different care objectives drive the datafication of care in the City’s data practices, the tensions that arise, and how workers engage in heuristic decision-making to work around the tensions.

\subsubsection{The matching objective. \hspace{0.3cm}} This objective was centered on serving the immediate needs of clients - collecting enough client information to match them with the resources they need. As seen in the yellow and orange boxes in Figure \ref{fig:system}, varying degrees of client information is collected at various points of the homelessness support system to provide clients with the right support. For example, the Call Center is seen as the front door to accessing homelessness support services. As P888 at the Call Center explains, \textit{"Our main goal here at [Call Center] is to get them into the shelter. And for that, we need to be able to just see that they qualify for shelter"}. When clients call in seeking shelter services, workers conduct a client intake (i.e., asking for their name, DOB, race, gender, and accessibility needs), keeping questions to a minimum to collect just enough data to identify a shelter space that fits the client's profile. When a client first calls the Call Center, a worker creates a client profile on HMIS. For future calls, the client can provide their HMIS number or their name and date of birth to access their records.

Shelters, on the other hand, collect more information since the client's immediate need for emergency shelter is met, allowing workers to engage with clients to apply for housing and other supports to ensure homelessness is brief and non-recurring. This also allows the workers to physically observe and interact with clients in a fixed location. A worker who has worked at shelters and the Call Center shared:

\begin{myquote}
\textit{"shelters would be different because you have eyes on people, you have counselors with them…. in my experience... they [clients] can [say] whatever the people want to tell you over the phone... but until you can really see them, see how they operate and how they behave then that's the true tell, so it's very different"}
\end{myquote}

Therefore, when clients first enter a shelter, workers conduct a more detailed intake where they ask clients about the specific supports they need, i.e., relating to their health, harm-reduction, accessibility, legal, and immigration needs. Workers assess client needs based on their responses and observations and triage the client as low, medium, or high-needs so the staff are aware of the level of support required. The intention of triaging a client at intake is to identify the supports the clients need and quickly provide them with resources. Throughout the client's stay, workers are trained to continually update the HMIS intake form as clients share new information with workers.

\subsubsection{The client’s privacy rights and agency objective. \hspace{0.3cm}} This objective was focused on respecting a client's agency and rights. A notable example is how client information is shared between different service providers. The green boxes in Figure \ref{fig:system} show the level of client information different groups can see. Even though different service providers use the same HMIS, the Call Center can see most of the client's history of services they received within the City, while other groups, such as the Streets Outreach group and shelters, can only see the client's service history within their organization. This is also why similar questions asked at the Call Center intake are asked at the shelter-level intake as well. Many shelter staff stated they agreed with this approach. P9277 said, \textit{"You have the people's privacy and confidentiality, it is very important... So, if we just spread their information right around, you know, and people can start saying you know what? I just read their file; I don't want to do that intake here, I don't need that, I don't need that hassle... we're not going to bring that person here"}. P9277 continued to explain that this approach helps clients get \textit{"a fresh start. You come here. We're not taking any of that stuff from before. Starting fresh here"}. Due to data sharing restrictions, there are no mechanisms for workers to check whether clients provide consistent responses across different service provider groups. And workers are trained to record client information as it is given to them to respect the client's agency.

\subsubsection{The equity objective. \hspace{0.3cm}}
This objective was centered around collecting data on client identities (\textit{"Equity objective"}). On a macro-level, this objective follows Canadian federal government mandates, which require communities to use data to determine which population groups should be prioritized for housing (e.g., those experiencing chronic homelessness, Indigenous people, seniors, black people, women, 2SLGBTQ+ people) \cite{reachinghomesdirectives}. The objective also helps the City keep track of how they are meeting their homelessness reduction targets for different population groups; understand how different client groups experience homelessness within the system to improve their services; and on an individual-level, when clients identify as one of the prioritized equity-deserving population groups at intake, the clients can be placed on prioritized waiting lists for housing services \cite{coordinated_toronto}. Questions about the client's race, pronouns, and Indigenous status are asked at every intake, at the Call Center, shelter-level, and by Streets Outreach groups. Asking equity-related questions can, however, introduce areas of frustration for clients. A common theme that emerged from conversations with frontline staff was that questions around a client's identity are one of the questions that workers face considerable pushback on. P888 explains, \textit{"when clients are coming to us, they're coming to a point where like, they've mostly exhausted everything. And they're now asking for help, like desperately asking like they're in crisis for the most part… Some are like, why are you asking me this information? Why don't I just get a shelter? I just need somewhere to sleep tonight."}

\subsubsection{Tensions between the matching objective and client's privacy rights and agency objective. \hspace{0.3cm}} Because different service providers cannot see a client's history from other service providers (\textit{Client's privacy rights and agency objective}), the same intake questions can be repeatedly asked \textit{across} service providers to identify the client's needs (\textit{Matching objective}). Clients can be asked the same questions \textit{within} the same service group, too (\textit{Matching objective}). For example, at the Call Center, due to the shortage of shelter availability in the City, clients rarely get a shelter bed on their first call and need to call the group again. Even if the client has done an intake and their client information is saved on HMIS, Call Center workers are required to ask some of the same intake questions if they call 24 hours later. P888 explains, \textit{"we have to do one intake per day after a 24-hour-period, because it resets so we have to ask like the main questions again, they don't necessarily have to go through their whole story, because it's already on file. But…we have to ask the same question multiple times."}. P888 continues, \textit{"because of the population that we serve, we do kind of have to be repetitive because things change so quickly. Like even for when we asked for a couple, we don't want to assume that the same couple who was together yesterday is together again today. Or if you had no mobility issues the day before, that you still have mobility issues the next day"}. Asking the same questions over and over can, however, add to the frustrations of the client with spillover effects into intake processes where clients do not want to answer questions or begin to provide inconsistent responses. When clients appear to provide inconsistent responses, workers are trained to respect the client's agency and record their responses as given (\textit{Client's privacy rights and agency objective}) posing tensions with the \textit{Matching objective}. 

\subsubsection{Tensions between the three objectives. \hspace{0.3cm}}
Questions around a client's identity, such as their pronouns, gender, or race, emerge as a contentious topic for some clients. On a macro-level, understanding client's identities can help the City track the types of clients entering the system and their outcomes (\textit{Equity objective}). At the street level, staff acknowledge that asking clients for their pronouns or gender is important as clients can be assigned to shelter/housing services reserved for the population group (\textit{Matching objective}) and also create an inclusive space for the client (\textit{Client’s privacy right and agency objective}). P35410 explained \textit{"it also helps us to direct the client to know what kind of gender they want to be addressed …I was doing the intake, I kept on using the wrong pronoun. And then the client might get agitated."} At the same time, asking these questions across different service providers can frustrate clients and cause them to provide inconsistent responses. P35410 recounted: 

\begin{myquote}
\textit{"When you can ask the client and you know…the client is already angry or let's say you see a client, he's already frustrated coming in because he was not getting the service or whatever it may be. And now he's able to get it, and then you ask the client - are you a male or female?... Question like, what race best describes you? The client who's upset? … sometime I've asked the client what race best describes you. He was white, and [he] said I'm black"}
\end{myquote}

Recognizing that workers should not assume to know the client's identity, workers are trained in such scenarios to record what the client tells them and to respect the agency of the client (\textit{Client rights and agency objective}). However, this approach also introduces challenges for the City’s \textit{Equity objective} because the HMIS database can show that a client identifies as multiple genders or races depending on the different sites the client was asked the question.

\subsubsection{Working around care-driven objectives through heuristic decision-making. \hspace{0.3cm}}

The data tensions that emerge between care-driven data objectives highlight the fundamental challenges in collecting unhoused client data and using data to inform policy and build decision-making tools in homelessness. As the City pursues different dimensions of care to protect their clients' rights, we find the City's original goal to minimize repeated client data collection is challenged. To ensure client privacy rights and due to the nature of experiencing homelessness, even if HMIS saves client information, different service providers cannot view complete client information and must repeatedly ask the same questions about accessibility needs, gender, and race. When repetitive information gathering is compounded by on-the-ground realities, i.e., severe shelter and housing shortages, we find existing client mistrust and frustration towards workers and the homelessness system is exacerbated - negatively impacting the clients' willingness to share accurate and consistent information with the City, which in turn impacts the worker's ability to find services the client may be eligible for \cite{Kryda_Compton_2009}. Moreover, questions from clients following the \textit{Equity objective} highlight a dilemma the City faces as it seeks to provide client care while respecting client agency amidst stark shelter and housing shortages. Notably, workers expressed facing pushback and inconsistent responses when asking clients about their race, gender, and pronouns. Clients who identify as part of equity-deserving groups can be prioritized for certain shelter/housing services, so sharing this information may have direct benefits. However, frontline workers are also trained to record what clients tell them to respect their agency. In these ways, we find care-driven data tensions emerge as the City’s intended data practices face on-the-ground realities of providing homelessness support.

Because workers know there are ongoing opportunities to gather client information during their stay within the homelessness system, we found workers attempted to resolve the intrinsic tensions by asking which data objective should take precedence to provide immediate support to clients. For example, during a shelter intake, the first author learned that a client had come to the shelter directly from a hospital, and workers observed the client was feeling weak and experiencing difficulties coming down the stairs. When the client sat down with a worker to do a shelter intake, they stated that they have no history/current medical issues. In this case, the workers recorded on the intake form that the client has no medical issues, explaining to the first author that they need to respect the client's agency, and the housing worker can delve into this further as they build rapport with the client. The worker privileged the client's \textit{`Privacy rights and agency objective'} over the \textit{`Matching objective'} to complete the intake process so the client could rest on a bed.

\subsection{Spatial, technological, and staffing-related variables can create imbalances in client data collected across service provider sites} \label{sec:physical}

Across all sites, our interview participants shared that collecting accurate client data during intake is critical because it helps inform workers on how to support them better. Collecting client intake data is also important because, following federal government guidelines, the City must maintain a real-time list of clients accessing services to determine which equity-deserving populations are overrepresented in the homelessness system and should be prioritized for housing and other related support \cite{reachinghomesdirectives, coordinated_toronto}. \textbf{While we found the care-driven data practices outlined in Section \ref{sec:r2} were intended to support standardized `care’ for clients, our visits to different shelter locations revealed that due to the variety of spatial, technological, and staffing-related differences across sites within the homelessness system, there could be significant differences in how client information was recorded on HMIS and the information clients were willing to share with workers.} Below, we provide examples of how a single data collection event, the shelter intake process when a client first arrives at a shelter, can differ due to various factors.

\subsubsection{Spatial factors. \hspace{0.3cm}}
We found client intakes can occur at different types of spaces depending on the physical structure of the shelter space. We noted that of the six shelters the first author visited, two shelters conducted intakes in semi-private rooms where others could not overhear conversations between the client and worker. At other locations, we saw intakes took place in crowded spaces, including in a dining hall and open areas by the shelter entrance. In one location, a worker would sit in an office while the client sat in the lobby and they would conduct the intake over a phone. While similar intake questions were asked at all locations, the first author noted that conducting intakes in loud settings made it difficult for the client and worker to hear each other, resulting in the loss of information. Moreover, a shelter worker explained that clients are often unwilling to open up when there are other clients or workers in the same space, especially when they are asked to disclose sensitive information. 

\subsubsection{Technological factors. \hspace{0.3cm}} \label{sec:r_tech}
There were also differences in the medium workers used to conduct intakes. Four of the six shelters we visited conducted intakes on the computer. This allowed workers to enter client responses into their computers directly. In two locations, workers printed out intake forms, recorded client responses on paper, and then entered client responses onto HMIS on the computer afterward. Some workers at these locations mentioned that intakes were done on paper because there were no computers in spaces where they could do intakes. The first author observed that some information got missed when workers were typing up client responses from the paper form into the computer because the workers have to handle multiple tasks at once. In many instances, the first author observed multiple clients would approach a worker inputting intake information into HMIS asking to do laundry or for their medication etc. We also found that on some occasions, the paper intake form had not been updated to include all the questions that HMIS asks, meaning some fields were left empty because the worker did not have the chance to ask the questions. The first author also observed that a shelter run by a partner organization (but funded by the City) used their own case management software in addition to the City's HMIS. As a result, staff would enter fewer client details than other shelters on the City's HMIS and include more detailed client information in their own case management software.

\subsubsection{Staffing factors. \hspace{0.3cm}}
The first author observed six in-person intakes into shelters, all conducted by different workers. The author observed that all the workers had different ways of doing client intakes, even if they were from the same shelter. If the worker forgot to ask a certain question on the intake form, some workers went back to the client to ask the client the missed question, while others recorded that the client did not provide an answer. We also found that worker shift schedules impacted intakes. In one observation, we noted that because a worker's shift was ending and the worker had not finished the client intake, another worker took over. Some information was lost in the process because the latter worker was not present for the first part of the intake and could only go off the previous worker's notes. Lastly, there were also significant divergences in what workers included in comment fields in the intake form. On the intake form, shelter workers are asked to triage the client into low, medium, or high-needs based on their observations and client responses. Some workers included detailed comments for their decisions, and others did not, explaining to the first author that that is the job of the housing worker to figure out.

Despite the City's intentions to collect client information following standardized care-driven data objectives, this section shows spatial, technological, and staffing-related differences across service providers within the homelessness system can impact the accuracy/consistency of intake client data collected on HMIS. In the next section, we show how the City and workers seek to work around these physical or resource-related differences by treating its client intake and, more broadly, assessment process as a continuous process that is integrated within each service provider's existing organizational practices all to prioritize providing direct care for clients.

\subsection{Client assessments are integrated within the City's data practices and start with the intake process} \label{sec:evolve}

Following the implementation of coordinated systems in the US and Canada \cite{reachinghomes, openingdoors}, homelessness support systems are mandated to use a common assessment tool for clients to assess their level of need. Many homelessness systems use standalone risk assessment tools such as the Vulnerability Index–Service Prioritization Decision Assistance Tool (VI-SPDAT) to meet this requirement \cite{Ecker2022}. Scores generated from such tools are often used to prioritize and match clients to different service types \cite{brown_2018} and prior works have raised a slew of concerns around its biased outcomes \cite{wholistic_2020, kuo23, Ecker2022, Cronley_2022, Slota_infra2022}. 

In stark divergence from many North American homelessness systems, when we asked frontline staff if they conducted client risk assessments similar to that of VI-SPDAT, workers explained that for the City, \textbf{the client assessment process was synonymous to needs assessment and were centered around treating client assessments as a continuous, relationship and trust-building process that is integrated within different service providers' existing staffing roles and client support relationships}. We learned the City had collaborated with Indigenous community organizations, anti-racism committees, frontline staff, service users, and those with lived experience of homelessness to develop this approach and purposely avoided using separate risk assessments like the VI-SPDAT because stakeholders found these tools can be biased.  By taking on this flexible approach, we found workers had strong buy-in towards the City's holistic approach to assessing and recording client information on HMIS. This approach also meant the City could overcome differences in how client information was recorded on HMIS due to staffing and spatial-related factors at various service providers (seen in Section \ref{sec:physical}). For example, at locations where there were comparatively fewer shelter workers compared to the number of shelter residents (where shelter workers do not have as much time to build rapport with clients), conducting client assessments was widely perceived to be the job of the housing worker, whose job was to meet with clients separately, build rapport with them, and continuously assess the client's risk. However, at locations with a greater shelter staff-to-client ratio, client assessments were viewed as the continuous documentation of the client's stay in the shelter program. \textbf{The City's fluid data practices also aligned with frontline workers' views that client data is an evolving and continuous construct that emerges as a byproduct of the care and rapport between clients and workers - especially as a City staff stressed - client needs are \textit{“never about the person but always an interaction between their environment and the person.”}}

Workers viewed the client intake process as the first stage of the client assessment process and the start of the relationship-building process between clients and workers. Depending on the staffing structure of different shelter sites, frontline workers perceived client assessments to be the primary responsibility of different workers. P9277, a shelter worker describes the first client intake as important as it marks the start of the relationship with the client:

\begin{myquote}
\textit{"It's very important because you get to be, first of all, you get to know the basics, who this person is, their age. They may not even always be telling you the truth, but that doesn't really matter. They're giving you something… that is that first point of contact, again to, you know, sort of feel each other out, right? If you do that, and you give them their room and you [client] get comfortable, it just builds that. I don't want to say trust because trust is very difficult to come by, particularly when you've, you know, been let down so many times so, but it's a first. It starts the beginnings of that working relationship"}
\end{myquote}

We also found that the City trained workers to update the client intake form whenever clients shared new/updated information with workers on HMIS. Because workers have no way to establish ground truths about a client’s circumstances unless they provide documentation (which clients are not always required to provide), treating client intake data as an evolving construct provided a workaround to collecting more accurate client data over time when clients are initially unwilling to share personal information (due to mistrust of the system, frustration of being asked the same questions, fatigue, hunger, and prior trauma). Taking this iterative data intake approach also meant HMIS could reflect any changes in the client’s circumstances. P9277 further explains the rationale for why client intake should not be a one-stop event:

\begin{myquote}
\textit{"I don't think intake should just stop at intake, right? I think it has to be revisited throughout because as you build that relationship, people may be more willing to release more information, right? So, for instance, if somebody comes in and you ask them next of kin, yes, I don't want to say anything. I don't want to give that information. And maybe a few months down the line, you [client] will say, you know what? Here's a good buddy of mine that you can contact or a family…You can always go in, you can go into intake, and you could click on update and update it...[HMIS] is always online it's always ongoing. So, you can always add information into [HMIS] with the changes"}
\end{myquote}

In sum, our study found that workers viewed client data on HMIS as an evolving construct that emerges through care and rapport built between clients and workers. And in face of data tensions and physical or resource-related differences in service providers, the City adopted a continuous, rapport-focused client assessment process that is integrated within existing service provider organizational structures to better meet client needs and follow federal data mandates.

\section{Discussion}

\subsection{Contributions to SIGCHI research on public sector data and care work (RQ1)}
Recent SIGCHI scholarship has extensively studied how public sector workers engage in discretionary care work, unpacking how workers translate client information into credible data and mediate complex dynamics between the realities of providing client care while adhering to formal data or algorithm requirements through interview studies \cite{bureaucraticjob_2021, saxena2021framework2, chi23paper, nielsen2023, Tracey_Garcia_2024_intermediation, kim_equity24, Tracey_Garcia_2024_automation}. Through an in-depth ethnography of a large homelessness system in Canada, our work extends and adds insights to these prior works by providing the following contributions. First, our work highlights the situated challenges of implementing neoliberal governance structures such as coordinated systems \cite{jonston_neoliberal17} and street-level frontline workers' discretionary data practices that emerge in response to these systems \cite{alkhatib_streetlevel19}. Through interviews and observations of frontline staff at multiple critical points of a large homelessness system, we show how clients' mistrust of the system combined with the different functional roles of various service providers and physical differences in service provider organizations can give rise to 1) care-driven data tensions (Section \ref{sec:r2}) and 2) inconsistencies in what information clients choose to share and how it is recorded on HMIS (Section \ref{sec:physical}). So while maintaining a centralized HMIS, in theory, should reduce repeated information gathering, and aggregated data should help drive data-driven policy insights, we find that in efforts to serve multiple dimensions of care and when faced with resource constraints, clients are asked the same questions at multiple points of their journey within the homelessness system, engendering client frustration and impacting the consistency of client data going into HMIS. Second, our findings identify how homelessness-specific factors shape frontline worker data practices in ways that are different from data and care work in other public sector domains. Notably, we learned that due to multiple reasons, including frustration, mistrust of the homelessness system, fatigue, prior trauma, as well as the rapidly changing circumstances of a client, information clients provide to frontline workers for HMIS entry can inevitably be inconsistent across the different sites they are asked the information. Moreover, because homelessness is largely caused by a systemic resource shortage issue (i.e., lack of affordable housing and shelter) and homelessness systems aim to provide low-barrier access to services \cite{reachinghomesdirectives}, we found clients are not obligated nor necessarily rewarded for cooperating with workers. Even though workers and the City argue collecting client data helps better match clients to available services, in the face of stark shelter and housing shortages, providing consistent answers for workers at intake did not necessarily increase the client's chance of getting a shelter bed or housing. These findings are different from data work in other domains, such as child welfare or asylum casework, where cooperation with caseworkers is sometimes found to increase the client's chances for a successful outcome (i.e., increased chances of bio-parents being reunified with their child or building a stronger asylum application) \cite{saxena2022chilbw, nielsen2023}. Following our study's research contributions, we discuss the higher-level implications of our study in relation to our RQs below.

\subsection{Data practices in homelessness systems and its implications for AI (RQ2)} \label{sec:disc1}

When workers are trained in the City's homelessness support data practices, workers learn the primary goal of collecting client information on HMIS is so that frontline staff can support clients. The secondary goal is so that the City can use the collected data for long-term service planning \cite{reachinghomesdirectives}. Our findings suggest frontline staff do not anticipate or plan on how data in HMIS can/will be used for the latter goal, data analysis (nor are they required to). There are information asymmetries, with frontline workers not knowing why or how data practices can constrain and impact the validity of data analysis methodologies. As an example, when the first author presented the purpose of the study at a service provider meeting to frontline staff early in the study to recruit participants, some workers challenged the author on why one needs to study data practices when the processes a client follows through the homelessness system is standardized. In a similar fashion, researchers and developers of AI models who use HMIS data for analysis are often trained to optimize model development but lack training in making sense of the data that arises from the social, physical, political, and organizational \cite{sambasivan21,dcai, boone2023, pine2015, vertesi2011}. 

These information asymmetries between frontline staff and users of HMIS for data analysis can, at the most basic level, limit the validity of data analysis findings. The ramifications of these information asymmetries, however, become increasingly problematic with the possibility of creating data cascades, that is, the negative compounding downstream effects of applying AI and ML techniques that ignore the data work \cite{sambasivan21} as researchers increasingly turn to developing AI models for homelessness.  Prior systematic literature reviews have found there has been growing interest in designing algorithms for homelessness, often using easily quantifiable but also deeply personal information such as a client's demographic and health-related information (e.g., physical, mental, or substance-abuse) \cite{showkat23, moon24}. Furthermore, \textbf{many cities in North America are increasingly using AI models using HMIS data or an amalgamation of public data which includes HMIS data, to predict an individual's risk of homelessness \cite{vanberlo2020, kuo23, lapolicylab, showkat23}. The adoption of these AI tools can create significant data cascades \cite{sambasivan21} because there are fundamental divergences in the values promoted in AI/ML research and values that promote human-centered models.}  Birhane et al. \cite{birhane22} and Showkat et al. \cite{showkat23} have found dominant values in AI/ML models focus on increasing the model's performance, generalizability, and novelty. Moreover, Alkhatib and Berstein \cite{alkhatib_streetlevel19} argue that street-level AI/ML models cannot exercise reflexive discretion when new situations or corner cases emerge as AI outcomes can only be fixed after the model produces an output. These vaunted AI/ML research values and intrinsic rigidities of street-level ML models are at direct odds with values emphasized in providing homelessness support. Much like many other areas of public sector work such as in child welfare, unemployment services, and refugee applications \cite{saxena2021framework2, kawakami2022, nielsen2023, bureaucraticjob_2021}, our findings show discretion and flexibility are key to frontline work for homelessness support and frontline staff value being able to apply data practices in ways that respond to the specific calls of the local community and stakeholders over generalizability (Section \ref{sec:evolve}).

Our findings in Section \ref{sec:r2} showed that operationalizing different human-centered values involves juggling competing care-driven objectives and resource constraints, which can create value tensions and yield mixed outcomes for stakeholders. For example, our findings from Section \ref{sec:r2} found that workers at different points of the homelessness system must repeatedly ask the same questions, and clients can provide inconsistent answers across sites out of frustration. Researchers using HMIS data for model development would thus need to determine which client's information to use as the ground truth. Even if researchers attempt to work around this issue by drawing on client records from multiple public databases \cite{kithulgoda_predictive_2022, eubanks2018automating, toros18}, the same challenges remain. Data cascades can emerge here through what Muller and Strohmayer term as `non-reversible forgetting practices of data science' \cite{muller_forgetting22}. Because data science workers in ML/AI frequently work in layered stages (e.g., data collection, curation, feature engineering, model training, deployment), AI/ML developers may forget or ignore the initial stage of the model development pipeline - in our case, the multiple interpretive actions frontline workers took to record HMIS data as they grapple with uncertainties that inherently underpin client data. Considering the above, we argue there is a critical need for researchers to exercise a deep understanding of the rationale and implementation of data work surrounding data collected on HMIS to ask if we need technical interventions for homelessness support \cite{baumer11} and if so, in what form.

\subsection{Rethinking risk assessments from deficit-based sorting models to holistic, relationship building approaches (RQ2)}
Since 2012 in the US and 2022 in Canada, homelessness systems in different jurisdictions have federally mandated the implementation of standardized assessments to gather client information accessing homelessness services \cite{reachinghomes, openingdoors, Ecker2022}. Accordingly, many communities employ risk assessment tools such as the VI-SPDAT or predictive risk models built using machine learning techniques to produce a vulnerability score for the client and accordingly prioritize them into different housing options \cite{Ecker2022, showkat23, eubanks2018automating, moon24}. Despite the widely studied limitations of these risk models, many communities continue to use the tools because of what Lu \cite{lu24betterthan} describes as a "better than nothing" sentiment \cite{Slota_infra2022, lechervarner} - there are no replacements. In our work, we were surprised to learn that the City diverged significantly from the abovementioned modular and static risk assessment approach adopted in other communities. \textbf{As outlined in Section \ref{sec:evolve}, we found that the City adopted a more holistic and iterative client assessment approach that emphasizes building rapport and trust with clients before they are asked personal and sensitive questions. }

\textbf{Interestingly, the City's data practices for client support through the homelessness system follow many of the design recommendations pitched by HCI researchers.} Saxena et al. \cite{saxena2021framework2} previously called on researchers to explore building holistic algorithmic assessments that can allow for heuristic decision-making. As highlighted in our Section \ref{sec:evolve} findings, the City's client assessments are not singular events; instead, it is a continuous process that occurs throughout a client's journey within the City's homelessness support system. Stapleton et al. \cite{stapleton22} also called on researchers to move away from predictive models. Our findings show that the City adopts these approaches. \textbf{Training material for the City's data practices instruct workers to adopt a trauma-informed and person-centered approach, emphasizing that the \textit{process} is more important than filling out forms; the immediate purpose of client assessments is not to `sort' clients into vulnerability brackets (and prioritize into housing) but rather to identify ways to `support' clients \cite{training_smis}.} Moreover, citing poor buy-in from workers that predictive risk assessments can improve their client's outcomes, prior SIGCHI work on this topic has stressed the importance of incorporating stakeholder perspectives, values, and a human-centered design lens \cite{saxena2021framework2, kawakami2022, hcai, vsd, muller2009participatory, stapleton22}. Interestingly, as seen in our findings from Section \ref{sec:evolve}, we found strong buy-in from workers on the City's data-driven client assessment approach, possibly because the City's data practices were collaboratively developed with community entities and incorporated feedback from stakeholders. All but one worker expressed concern about the amount of client data being stored on HMIS. Most workers stated they were comfortable with client information being collected because they only record information that a client is willing and has consented to share; the City has well-established information-sharing policies, so no client information is shared with external parties unless they have explicit consent from clients; and client information is collected to \textit{support} clients. At the same time, the City's data practices also fundamentally differ from many HCI researchers' algorithmic design guidelines, which do not question whether using AI tools for homelessness is appropriate in the first place and instead focus on proposing guidelines on improving human-AI partnerships \cite{kuo23, kawakami2022}.

While the City's data practices provide an exemplar case for holistic decision-making for the public domain, it is also important to note the current limitations of the City's approach to client assessment and support. Our findings from Section \ref{sec:r2} and \ref{sec:physical} highlight there are limitations in how the City's \textit{intended} data practices are operationalized on the ground \cite{mothilal24}. \textbf{The City's approach to client assessment places a large burden on frontline workers to take up the rapport component of care work (Section \ref{sec:evolve}) \cite{karusala19, nielsen2023, Slota_infra2022}.} However, this can be challenging in the face of resource constraints and individual differences. Through visits to different shelters and conversations with shelter workers, we learned workers have different ways of engaging with clients and documenting client interactions (Section \ref{sec:physical}).  Shelters can vary widely, with some having more staff, on-site support services (such as doctors), and recreational programming that facilitates rapport-building with clients. As a result, there can be many variations in how client assessments are carried out. Moreover, in the face of shelter and permanent housing shortages, systemic resource constraints limit the worker's ability to support clients \cite{eubanks2018automating, Desmond_2016}.

Our findings on the City’s client assessment approach carry two high-level implications for research on risk assessments. First, we showcase how holistic client assessments can be designed and implemented for decision-making over deficit-focused risk assessments. Second, we problematize current AI decision-support tools in homelessness that assess client risk to prioritize clients to services \cite{showkat23, moon24}. Our findings show that client needs evolve depending on what information a client is willing to share with workers and dynamic interactions between the person and their environment (Section \ref{sec:evolve}). AI developers interested in designing tools for homeless systems must, therefore, understand and account for these temporal factors lest they incur unintended harm.

\subsection{Implications for future computational HCI research on homelessness support systems and beyond (RQ3)}

In this section, we reflect on our study findings and outline future paths for how HCI researchers can support high-stakes public sector domains such as homelessness and their frontline staff.

We reiterate the oft-repeated argument that HCI researchers should deeply engage with stakeholders and communities to understand, on the ground level, their needs and perceptions around data and technology \cite{aragon2022human}. In our work, we found the City's approach to assessing clients – as a continuous and rapport-focused exercise between workers and clients – had strong buy-in by frontline staff \textit{because} the City had developed their data practices through extensive collaborations with community organizations and impacted stakeholders. We thus encourage HCI researchers who strive to design technologies for homelessness to consider the fundamental tenet that technical interventions are developed within the context of systemic constraints, organizational processes, and asymmetrical power dynamics between workers and clients \cite{saxena22, kuo23, selbst2019fairness, kelly23, Kawakami_2024, barabas20, saxena2023algorithmic}.

Our study highlights the critical role frontline workers play in mitigating client mistrust and lack of confidence in the homelessness system to carry out the City’s intended data practices. As homelessness support systems increasingly adopt AI/ML decision-making tools, we argue that the role of frontline workers as primary client advocates will increase further. Accordingly, HCI researchers should explore ways to empower workers to understand how client data is used to generate AI/ML-based decisions and how they may successfully contest its outcomes \cite{showkat23, karusala_contestability24}. Recent work has already found frontline workers engage in critical discretionary intermediation practices, translating complex client circumstances into structured data entries in HMIS and prioritization algorithms \cite{Tracey_Garcia_2024_intermediation}. Workers also act as the first line of defense for historically marginalized client groups and undocumented clients who may be unwilling to share their information due to fear of deportation or incarceration \cite{karusala_contestability24}. Yet, we find there are fundamental information asymmetries wherein frontline workers do not know the significance of their data work (Section \ref{sec:disc1}). Empowering frontline workers to understand how the data they collect is used in AI/ML modeling will assist them in better supporting their clients and pursuing recourse against AI/ML decisions that negatively impact their clients.

Our findings carry implications for scholarship on data and care work in other public sector services. Domains such as child welfare, higher education, welfare benefits, and job placement services share similarities with homelessness services wherein these domains are increasingly turning to AI/ML decision-making tools to categorize clients through a risk lens \cite{angwin, eubanks2018automating, kuo23, bureaucraticjob_2021, saxena2021framework2, chi23paper, kelly24}. Prior works find these systems treat client “risk” through a deficit-based lens; these tools aim to assess and minimize risk rather than improve client outcomes by focusing on their strengths. With its holistic, asset-based approach to homelessness support, our study exemplifies how alternative approaches to client care are both possible and desirable. We also find that in lieu of risk-focused prioritization tools, workers desire improvements in the technical tools they use, including improving user interfaces with customized drop-down selection options and the ability to see updated information on clients more quickly \cite{Holtenshifting_2020, gondimalla24}. Given the iterative and dynamic nature of providing client support in public services, we, therefore, encourage HCI researchers to explore how technologies can empower stakeholders over building predictive risk models.

In sum, frontline workers in contexts of homelessness are uniquely positioned as primary client advocates who must understand and contest how AI/ML decisions impact vulnerable people. Opportunities exist for HCI researchers to study "risk" as an evolving, holistic construct and guide decision-making by presenting a comprehensive view of client circumstances over time.

\section{Limitations}
Since adopting the coordinated systems in the US and Canada, different communities have had considerable latitude in how they implement data-driven service delivery to clients \cite{Ecker2022}. Our study focused on one Canadian city that is experiencing a high demand for homelessness services. Other communities can be subject to different privacy and consent regulations and housing objectives that impact their data practices and work by frontline staff. Our findings may not be generalizable to other settings. Moreover, our study focused on frontline staff perspectives on the City's data practices. Our work did not interview individuals who may be experiencing or have lived experiences of homelessness, and they will likely have different views particularly around privacy and information sharing consent procedures.

\section{Conclusion}

We conducted an ethnographic study, interviewing and observing how frontline staff collect client information to study a homelessness system’s data practices for a large city in Canada. Through our work, we show the situated data practices of homelessness frontline staff, revealing how they engage in heuristic decision-making, prioritizing different care-driven data objectives over others to help clients access critical services they need. We also show that client data saved on HMIS is an evolving construct. These findings have implications for the growing interest in using AI to combat homelessness as poor understanding of data work can result in negative downstream effects for AI models built using this data \cite{sambasivan21}. Through our work, we encourage HCI researcher to move away from designing context-unaware models validated by posthoc tests on racial fairness and goodness of fit metrics \cite{sambasivan21, selbst2019fairness}. Instead, we encourage researchers to work on research questions that are of interest to stakeholders and validated by downstream users.

\begin{acks}
This research was supported by the NSERC Discovery Early Career Researcher Grant RGPIN-2022-04570, Schwartz Reisman Institute for Technology and Society Graduate Fellowship, and Carnegie Mellon University Presidential Postdoctoral Fellowship Program. Opinions, findings, and conclusions expressed in this paper are those of the authors and do not necessarily reflect the views of our research site partners. We sincerely thank our collaborators and study participants who have allowed us to conduct this ethnography as they work tirelessly to support their clients. Additionally, we thank our anonymous reviewers whose suggestions and comments helped improve this manuscript.

\end{acks}

\bibliographystyle{ACM-Reference-Format}
\bibliography{sample-base}



\end{document}